\documentclass[sigconf]{acmart}
\AtBeginDocument{%
  \providecommand\BibTeX{{%
    \normalfont B\kern-0.5em{\scshape i\kern-0.25em b}\kern-0.8em\TeX}}}

\setcopyright{acmlicensed}
\copyrightyear{2024}
\acmYear{2024}
\acmDOI{10.1145/3650105.3652298}

\acmConference[FORGE]{1st International Conference on AI foundation models and software engineering}{April 14--20,
  2024}{Lisbon, Portugal}

\acmBooktitle{FORGE '24: 1st International Conference on AI foundation models and software engineering,
 April 14, 2024, Lisbon, Portugal} 
\acmISBN{979-8-4007-0609-7/24/04}

\usepackage{graphicx}
\usepackage{array}
\usepackage{longtable}
\usepackage{multirow}
\usepackage[frozencache ,cachedir=.]{minted}
\usepackage{supertabular}
\usepackage{arydshln}
\usepackage{xcolor, xparse}
\usepackage{soul}

\usepackage{todonotes}
\begin{document}

\title{An Exploratory Investigation into Code License Infringements in Large Language Model Training Datasets}

\author{Jonathan Katzy}
\email{J.B.Katzy@TUDelft.nl}
\orcid{0009-0005-9574-2414}
\affiliation{%
  \institution{Delft University of Technology}
  \country{Delft, Netherlands}
}

\author{Răzvan-Mihai Popescu}
\email{R.Popescu-3@student.TUDelft.nl}
\orcid{0009-0003-6251-770X}
\affiliation{%
  \institution{Delft University of Technology}
  \country{Delft, Netherlands}
}

\author{Arie van Deursen}
\email{Arie.vanDeursen@TUDelft.nl}
\orcid{0000-0003-4850-3312}
\affiliation{%
  \institution{Delft University of Technology}
  \country{Delft, Netherlands}
}

\author{Maliheh Izadi}
\email{M.Izadi@TUDelft.nl}
\orcid{0000-0001-5093-5523}
\affiliation{%
  \institution{Delft University of Technology}
  \country{Delft, Netherlands}
}

\begin{abstract}
Does the training of large language models potentially infringe upon code licenses? Furthermore, are there any datasets available that can be safely used for training these models without violating such licenses?
In our study, we assess the current trends in the field and the importance of incorporating code into the training of large language models. Additionally, we examine publicly available datasets to see whether these models can be trained on them without the risk of legal issues in the future.
To accomplish this, we compiled a list of 53 large language models trained on file-level code. We then extracted their datasets and analyzed how much they overlap with a dataset we created, consisting exclusively of strong copyleft ~\mbox{code.}

Our analysis revealed that every dataset we examined contained license inconsistencies, despite being selected based on their associated repository licenses. We analyzed a total of 514 million code files, discovering 38 million exact duplicates present in our strong copyleft dataset. Additionally, we examined 171 million file-leading comments, identifying 16 million with strong copyleft licenses and another 11 million comments that discouraged copying without explicitly mentioning a license.
Based on the findings of our study, which highlights the pervasive issue of license inconsistencies in large language models trained on code, our recommendation for both researchers and the community is to prioritize the development and adoption of best practices for dataset creation and management.

\end{abstract}


\maketitle

\section{Introduction}
The datasets for training Large Language Models (LLMs) have expanded rapidly, mirroring the increase in the number of parameters in cutting-edge models. This surge has necessitated the quick creation of numerous large datasets for training purposes. Alongside this growth in model size, there has been a notable shift in adapting Programming Language Models (PLMs) for end-user applications. This shift has piqued the interest of businesses looking to utilize these models commercially, leading to rising concerns about the legal implications of using copyrighted data in such large-scale training datasets.
The significance of adopting permissive licenses in training LLMs has been recognized by entities like Together Computer~\cite{together2023redpajama} and The BigCode Project~\cite{kocetkov2022stack}. They have released models~\cite{li2023starcoder, allal2023santacoder} and datasets~\cite{kocetkov2022stack, together2023redpajama}, 
claiming that they consist exclusively of permissively licensed code.
Permissive licenses, like the MIT and Apache licenses, allow for minimal restrictions on software use, modification, and distribution, even allowing incorporation into proprietary software. In contrast, strong copyleft licenses, such as the GNU General Public License (GPL), require that any derivative works also be open source, maintaining the same user freedoms as the original software.

In similar domains, there have been legal cases centered on copyright holders objecting to the use of their data for training LLMs~\mbox{~\cite{gettyimages2023, nytimes2023, huckabee2023}.} The majority of these disputes involve claims of lost profits due to unlicensed data used in model training. Additionally, there are complaints about potential damage to a company's reputation when its name is linked to low-quality or inaccurate information.
Some companies have demanded the deletion of LLM weights trained using their data, a move that could cost the model creators millions of dollars~\cite{nytimes2023}. The common thread in these legal cases, and a looming concern for future LLM development, revolves around the extensive scraping of online data without regard for associated licenses or ownership rights. This practice mirrors the prevalent approach to code dataset compilation, where most data is scraped from platforms like GitHub without considering ~\mbox{licensing.}

One key challenge in the wider adoption of LLMs stems from their classification as Foundation Models (FMs), which are trained on extensive datasets and then fine-tuned for specific tasks. This practice of reusing weights, however, introduces risks for end-users. Concerns such as data memorization and membership inference attacks, as highlighted in recent studies~\cite{yang2023code, carlini2023quantifying,kaswan2023traces,al2023targeted,Hu2022Membership,yang2023gotcha}, enable the detection of copyrighted or licensed content in both the original and fine-tuned model versions. Additionally, distributing these models could be interpreted as redistributing copyrighted material. Therefore, it is essential to mitigate the risk of future legal challenges~\cite{al2023ab}.
To determine the potential for future legal challenges associated with LLMs, we adopt a bottom-up approach. The primary source of such legal issues is likely to be the data used for training these models. While some studies have focused on compiling 'safe' datasets, they often overlook the origin of their data. To confirm this concern, we conducted a comprehensive survey of the LLM field, collecting data on the models currently in use and their training datasets. We then analyzed this information to identify potential future licensing problems.
More concisely, we will be answering the following research questions:
\begin{enumerate}
    \item How has the interest in including source code in the training of both generic and specific language models evolved over time?
    \item What is the minimum level of existing strong copyleft-licensed code in the training data of PLMs?
    \item 
    What types of sensitive information might be present in the datasets of PLMs?
\end{enumerate}

Our contributions are as follows.
\begin{itemize}
    \item We provide a detailed overview of how source code is utilized as a data source in contemporary FMs,
    \item We compile a comprehensive summary of the datasets currently employed in training,
    \item We assess the exposure of FMs to potential issues with copyright and license holders, focusing on publicly available datasets,
    \item We introduce a dataset comprising the opening comments from 171 million code files, designed to aid in identifying copyright and licensing concerns in future research.
\end{itemize}

\section{Background}
We introduce this work with an overview of the literature. When it comes to potential legal challenges, we focus on the possibility of extracting verbatim copies of code from the training set. We show what works have focused on membership inference attacks (showing that a given code file is included in the training set of a model) and memorization (verbatim copying of training data in LLMs). Finally, we cover the bridge between the theoretical limitations of LLMs and the legal field, including possible safeguards that have been suggested in the literature when working with code in LLMs.

\paragraph{Membership Inference on Code}
Being able to confirm if a code file has been used in the training of a model is important information to determine whether a license could have been infringed. Determining whether data has been seen during the training of a model is known in the literature as a membership inference attack.

Beyond the scope of Large Language Models for Software Engineering, membership inference attacks have been widely studied in the machine learning literature~\cite{Hu2022Membership}. Traditionally, topics related to the privacy of people that are contained in datasets were addressed, although the main focus was on tasks such as classification~\cite{Hu2022Membership}.
Recently, researchers have started using membership inference attacks to extract training data from code models. As mentioned above, there are two distinct settings for membership inference for code models, representation-generating models, and output-generating ~\mbox{models.}

For the representation-generating code models, the BUZZER~\cite{zhang2023buzzer} approach was proposed. In this approach, the authors attempt to identify if a code fragment had been present in the pretraining data for the models; CodeBERT~\cite{feng2020codebert}, GraphCodeBert~\cite{guo2021graphcodebert}, Unixcoder~\cite{guo2022unixcoder}, and CodeT5+~\cite{wang2023codet5}. The authors used a white box approach (BUZZER had access to the internal states of the model), a gray box approach (BUZZER had access to the internal states of a shadow model), and a black box approach (BUZZER had no access to any internal states) approach, and showed that they achieved $90\%$ accuracy when determining whether code was used during training in the white box setting, around $80\%$ accuracy in the gray box setting and around $60\%$ accuracy in the black box setting~\cite{Hu2022Membership}.

In the output-generating approach, the Gotcha~\cite{yang2023gotcha} approach was proposed. In the Gotcha approach, the open-source CodeGPT~\cite{lu2021codexglue} model was analyzed to determine its exposure to data leakage. In this paper, the membership inference attack trains surrogate models, to mimic the behavior of CodeGPT, and later a classifier to determine whether data was or was not in the training data of CodeGPT. For the surrogate models, different architectures were used to determine the effectiveness of the attack, based on how much knowledge the attacker had of the target model. The evaluated models were: CodeGPT (when the architecture and training data are known), GPT-2 (only the architecture is known), Transformer (only the type of architecture is known), and LSTM (no architecture or training data is known). In the best-performing setting, the model is known and 20\% of the training data is used. Gotcha achieved an error rate of only 10\% and an AUC of 0.98\%. This high performance shows that code models are vulnerable to membership inference attacks when generating code, and together with the high accuracy reported by BUZZER they validate each other's results~\cite{yang2023gotcha}.

\paragraph{Memorization}
While the design of membership inference attacks is still very new when applied to Foundational Models for code, measuring and preventing a model from returning memorized code has had more attention in the literature. When looking at memorization in the output of Large Language Models, there are many similar definitions to what a memorized output is. The one thing they have in common is that they look for overlap between outputs from the model and substrings of the dataset. Some papers only look at exact duplicates of outputs~\cite{carlini2023quantifying}, while others look for close matches, and yet others will add constraints to how long a substring must be before it can count as memorization when output by a model~\cite{yang2023code}.

When dealing with the memorization of code, it has been shown that for models, the number of parameters correlates to the amount of memorization~\cite{kaswan2023traces, carlini2023quantifying}. As models have been increasing in size rapidly over the last years, memorization of training data (including code) will become an ever-increasing issue. While the largest models are often not open-sourced by the creators, some works look at the rate at which code is memorized by models when the train set is ~\mbox{available.}

It has been shown the $81\%$ of the top-100 outputs generated by StarCoder~\cite{li2023starcoder} are copied directly from GitHub, and $75\%$ of the outputs generated by InCoder \cite{fried2023incoder} are copied from GitHub. Furthermore, when evaluating the open-source model CodeParrot, they found that the repetition of code in the dataset, as well as querying a model multiple times, and for longer output sequences, increased the chance that a memorized code snippet was returned. Finally, the authors also manually identified a taxonomy of memorizations that code models output, where the most common type of memorization was returning the license information for a file~\cite{yang2023code}.

Although the detection of exact memorization seems like an easy task that can be prevented by a filter, Copilot, an implementation of the Codex~\cite{chen2021evaluating} model, can evade its filters~\cite{ippolito2022preventing}. This also adds an extra layer of confidence in the model's ability to generate original content, while not addressing the underlying issues with LLMs.

\paragraph{Relations to the Legal Field}
A final question is to what extent foundation models can claim that their outputs are protected under \emph{fair use}~\cite{henderson2023foundation}:
a substantial part of an output must be an exact copy of the original, for fair use to no longer be applicable. A brief analysis has been done on extracting examples of strong copyleft licensed code using the ChatGPT model (GPT-4), while also evaluating the average match percentage of the code-cushman-001, code-davinci-001, and code-davinci-002 models to be around $50\%$ when prompting them with function signatures of the Linux kernel. The authors of this paper continue to discuss the problems with language models that remove copyright information that must be copied when using code from a file~\cite{henderson2023foundation}.

As a solution to the issue of copyright infringements, the authors suggest a number of technical fixes.
First, they suggest that training data are selected based on the license that is assigned to a file, similar to how some datasets are created~\cite{kocetkov2022stack}. They also suggest to focus on the quality of the data being used. Suggesting to remove duplicate data, something that has been shown to increase memorization of models~\cite{yang2023code}. Furthermore, they suggest adding filters to the output of a model, which may be beneficial, however, it has been shown to give false confidence in the model~\cite{ippolito2022preventing}.

\section{Legal Aspects}
Having covered the limitations of LLMs, with respect to memorization and vulnerability to membership inference attacks, we next see how these limitations manifest themselves in the real world. We first address three lawsuits that are currently being litigated in courts concerning the data contained in models and datasets. Then we give an overview of common licenses that are applied to source code and what implications their requirements have for LLMs that may have been trained on them.

\subsection{Lawsuits}
To understand the qualms data owners have with their data being included in training data, we give a brief overview of the legal troubles surrounding the \textit{Books3} corpus, the lawsuits surrounding the stable diffusion models and how they can produce copyrighted imagery, and finally the lawsuit between the New York Times and OpenAI regarding the inclusion of New York Times articles in the training of their models.

\paragraph{Books3}
The \textit{Books3} corpus is a dataset of books scraped from an online site that distributed materials protected by copyright. Its original aim was to level the playing field between big AI companies and individuals working with Large Language models. After its creation, it was included in a dataset that combined existing datasets to create an extensive corpus and released under the name, ~\mbox{\textit{The Pile}~\cite{gao2020pile}.}

This dataset was eventually used by several companies such as EleutherAI, Bloomberg, and Microsoft to train for-profit large language models. In the United States of America, a lawsuit was launched by many authors (who have authored works included in \textit{Books3}) who demand that the mentioned companies stop using their books permanently (it is unclear if they want current models to be removed) along with compensation for the use of their ~\mbox{works~\cite{huckabee2023}.}

The main arguments brought forth in the lawsuit allege that Microsoft, Meta and Bloomberg: 
\begin{itemize}
    \item All developed LLMs while knowing that the training data was copyrighted.
    \item Did not attempt to obtain a license for the copyrighted works while knowing that the original works were obtained illegally.
    \item All chose to use stolen work to train models with the goal of generating a profit.
\end{itemize}

Furthermore, a Danish interest group, Rights Alliance, has issued a DMCA takedown request for the dataset \textit{The Pile} for containing the \textit{Books3} corpus, making the entire dataset unavailable for download~\cite{rightsalliance2023}.

\paragraph{Stable Diffusion}
In a related court case also running in the United States of America at the moment, Getty Images, a provider of digital images, claims that their copyright has been infringed by StableAIs' Diffusion model, claiming that StableAI used 12 million images curated by Getty Images in the training of their model~\cite{gettyimages2023}. The lawsuit launched by Getty Images is based on three main arguments:
\begin{itemize}
    \item The StableAI models sometimes produce images that contain the recognizable Getty Images banner, which would be an infringement of the trademark.
    \item The images used for training were scraped from the Getty Images site, which is against the license of their website.
    \item Getty Images was not contacted for a license of their images included in the dataset, which they spend a large amount of money curating, including captions, titles, keywords, and other metadata.
\end{itemize}

\begin{figure*}[ht]
    \centering
    \includegraphics[scale=.68]{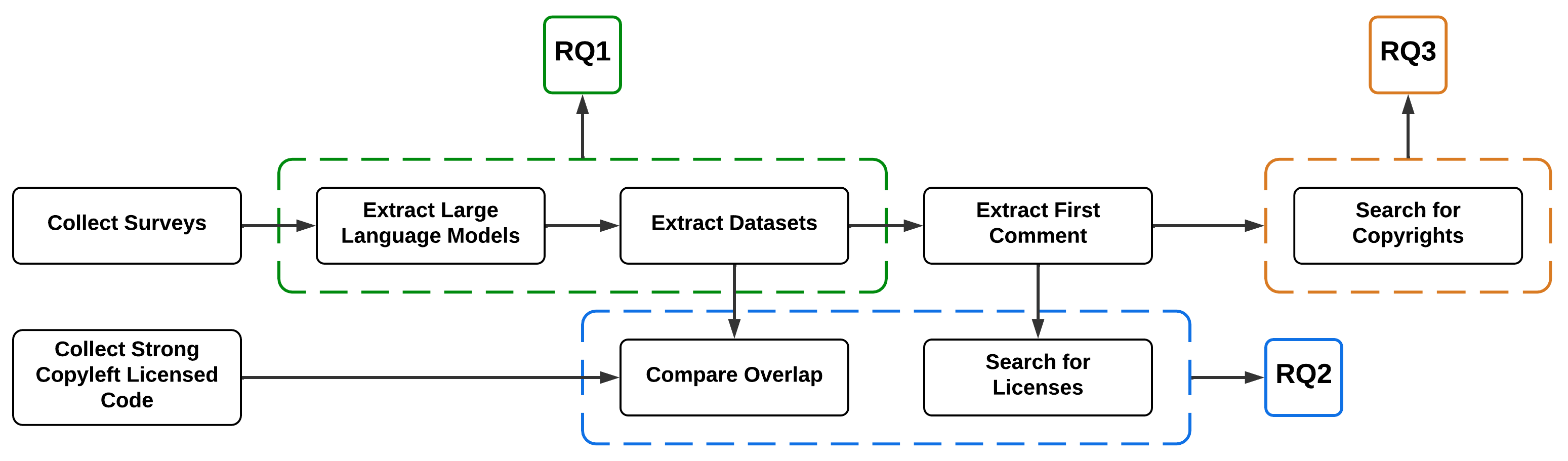}
    \caption{Graphical overview of the presented approach}
    \label{fig:overview}
\end{figure*}
\paragraph{OpenAI}
The most recent and relevant lawsuit currently in progress is between The New York Times, Microsoft, and OpenAI. The lawsuit is based on the usage of articles written for The New York Times and copyrighted by The New York Times in the training of LLMs distributed by OpenAI and Microsoft~\cite{nytimes2023}. 
Similarly to the previous two lawsuits, the issue is centered on the inclusion of copyrighted material in the dataset used to train AI models. The issues that The New York Times has raised in this lawsuit can be summarized as follows:
\begin{itemize}
    \item The models are trained on copyrighted text that The New York Times invested money into creating, without having obtained a license for its use.
    \item Models that output articles on the same subject as New York Times articles reduce the number of people who might subscribe to The New York Times.
    \item The models sometimes hallucinate incorrect 'facts', but bring it in connection with the New York Times, which harms the image of the New York Times.
    \item The models can output verbatim copies of New York Times articles to people who do not have a subscription.
\end{itemize}

Although the first two points are similar to the issues raised by Getty Images and the \textit{Books3} corpus, the New York Times lawsuit goes more in-depth into the technical limitations of current LLMs and how they affect suppliers of training data.

The first point, of AI models hallucinating 'facts', is backed up by screenshots of prompts in the lawsuit. The lawsuit focuses on Bing Chat, which when prompted for specific parts of an article produces text that is not from a New York Times article. Furthermore, the lawsuit shows evidence of Bing Chat creating citations of people that are supposedly in a New York Times article, while this was not the case. Finally, the last example of hallucinations is Bing Chat claiming facts are published in New York Times articles when the articles it references were never published.

The second point, where AI models copy data from the training set, is backed up in the lawsuit filings by showing examples of outputs from the GPT-4 model and comparing it to the text published in the New York Times, which was largely identical. Furthermore, they showed examples of GPT-4 returning exact copies of New York Times articles when they prompted it by saying they were blocked by a paywall.

In contrast to the previous lawsuits, there had been rounds of negotiations between The New York Times and OpenAI about licensing the data from their articles; however, they did not come to an agreement. The defense of OpenAI to the allegations of copyright infringement is that the use of articles in the training of models can be seen as transformative and should be allowed under fair use. Furthermore, they argue that the methods used by The New York Times to extract their articles from the GPT-4 model are against OpenAI terms of service and are not permitted to be used.

Finally, The New York Times not only sues for damages they perceive to have been inflicted on them by Microsoft and OpenAI, but they also sue for the destruction of all GPT or LLMs weights that were trained on datasets containing New York Times articles, as well as the destruction of all datasets containing their articles.

\subsection{Existing Licenses}
The code used in the training and fine-tuning of LLMs is governed by an extensive range of licenses that impose various restrictions on the use and distribution of the software. While describing any potential limitations and restrictions, each license category gives users a specific set of rights. 
Licenses can be classified into three main groups: strong copyleft, weak copyleft, and permissive.

\begin{table*}[h]
    \centering
    \caption{Literature Surveys used to identify Large Language Models}
    \begin{tabular}{p{120mm}|>{\centering\arraybackslash}p{15mm}}
    \toprule 
      \multicolumn{1}{c|}{\textbf{Title}} & \multicolumn{1}{c}{\textbf{Reference}}\\
       \hline
       Software Testing with Large Language Model: Survey, Landscape, and Vision & \cite{wang2023software}\\
       A bibliometric review of large language models research from 2017 to 2023 & \cite{fan2023bibliometric}\\
       A survey of large language models & \cite{zhao2023survey}\\
       Large Language Models for Software Engineering: A Systematic Literature Review & \cite{hou2023large}\\
       Large Language Models Meet NL2Code: A Survey & \cite{zan2023large}\\
       A Survey on Large Language Models for Software Engineering & \cite{fan2023large}\\
       \bottomrule
    \end{tabular}
    \label{tab:literature_surveys}
\end{table*}

\begin{table*}[t]
\centering
\caption{Division of Large Language Models based on training data}
\label{tab:models} 
\begin{tabular}{p{3cm}|p{3cm}|p{3cm}|p{3cm}}
\toprule
 \multicolumn{1}{p{2.25cm}|}{\textbf{Only Natural \newline Language}} 
 & \multicolumn{1}{p{2.25cm}|}{\textbf{Method-Level Code}} 
 & \multicolumn{1}{p{2.25cm}|}{\textbf{Permissive \newline File-Level Code}} 
 & \multicolumn{1}{p{2.4cm}}{\textbf{Non-permissive File-Level Code}} \\
\midrule 
GPT-3 & PLBART  & CodeGen\textsuperscript{1,2,9} & Codex\textsuperscript{15} \\
T5 & Tk-Instruct & InCoder\textsuperscript{15} & CodeT5\textsuperscript{1} \\
BART & ERNIE-Code & FLAN-T5\textsuperscript{3} & BLOOM\textsuperscript{1,7} \\
mT5 & PyMT5 & LLaMa\textsuperscript{1} & Galactica\textsuperscript{15} \\
CPM-2 & LaMDA & CodeGen 2\textsuperscript{2,3,9} & Baichuan 2\textsuperscript{15} \\
PanGu-$\alpha$ & InstructGPT & StarCoder\textsuperscript{3} & QWEN\textsuperscript{15} \\
T0 & CodeBERT & Gopher\textsuperscript{10} & Skywork\textsuperscript{8} \\
UL2 & CodeRetriever & CodeT5Mix\textsuperscript{11} & Pythia\textsuperscript{2} \\
OPT & TraceBERT & CodeRL\textsuperscript{11} & Jurassic-1\textsuperscript{2} \\
NLLB & GraphCodeBERT & AlphaCode\textsuperscript{15} & JuPyT5\textsuperscript{15} \\
GLM & BERT Overflow & PaLM\textsuperscript{6} & MT-NLG\textsuperscript{2} \\
FLM & CoText & LLaMa 2\textsuperscript{1} & PyCodeGPT\textsuperscript{15} \\
GShard & PanGu-Coder & WizardLM\textsuperscript{1} & U-PaLM\textsuperscript{6} \\
HyperClova & CodeGPT & CodeT5+\textsuperscript{11} & PanGu-$\Sigma$\textsuperscript{15} \\
Yuan 1.0 & CodeGPT-adapted & WizardCoder\textsuperscript{3} & PaLM 2\textsuperscript{15} \\
GLaM & CoditT5 & SantaCoder\textsuperscript{3} & Mistral\textsuperscript{15} \\
AlexaTM & SPT-Code & PaLM-Coder\textsuperscript{6,13} & GPT-C\textsuperscript{15} \\
WeLM & FLAN & Vicuna\textsuperscript{1} & PolyCoder\textsuperscript{15} \\
BERT & UnixCoder &  Stable Code\textsuperscript{3} & GPT-3.5\textsuperscript{15} \\
mBART & PanGu-Coder-FT & StableLM\textsuperscript{2,3,4} & Code LLaMa\textsuperscript{1,14} \\
GPT-1 & PanGu-Coder 2 & StableLM Zephyr\textsuperscript{2,3,4} & GPT-NeoX\textsuperscript{2} \\
XLNet & T5-Learning & Japanese StableLM\textsuperscript{4} & CodeGeeX\textsuperscript{2,5,15} \\
Sparrow &  & Stable Beluga\textsuperscript{1} & CodeParrot\textsuperscript{5} \\
PRCBERT & & & GPT-CC\textsuperscript{12} \\
seBERT & & & Chinchilla\textsuperscript{15} \\
ALBERT & & & GPT-J\textsuperscript{2} \\
RoBERTa & & & FIM\textsuperscript{15} \\
OPT-IML & & & GPT-Neo\textsuperscript{2} \\
ERNIE 3.0 & & & Falcon\textsuperscript{2} \\
GPT-2 & & & CuBERT\textsuperscript{1}\\
WebGPT & & & \\
\bottomrule
 \end{tabular}
\end{table*}
\paragraph{Permissive}
This category of licenses enables users to utilize, alter, and redistribute the software with great freedom and without being subject to strict limitations. The non-restrictive nature of these licenses allows for flexibility in integrating code into both proprietary and open-source projects. This characteristic stems from their ability to support changes and the production of derivative works without the requirement of sharing under an identical license \cite{apache-license}. Permissive licenses enable the incorporation of code with permissive licenses into projects that have varying licensing needs. Users have the privilege of using the altered code under a separate license or retain it as proprietary. In contrast to copyleft licenses, permissive licenses only require users to provide attribution and exempt the author from liability. 
\paragraph{Weak Copyleft}
As their name suggests, weak copyleft licenses allow the integration of code into proprietary projects without requiring the entire derived work to be open-source \cite{gnu-copyleft}. In other words, only modified parts of the original code must be released under the same weak copyleft license. This category of licenses achieves a middle ground between proprietary software and collaborative open-source development. Aside from sharing modified code, weak copyleft licenses also mandate attribution to the original authors, and potentially a disclaimer of liability. 
\paragraph{Strong Copyleft}
What sets strong copyleft licenses apart from weak ones is their enforcement of a reciprocal condition that any derivative work that incorporates or alters the original code must be distributed under the same strong copyleft license. This share-alike condition ensures that the same rights are preserved in subsequent versions and derivatives of the software \cite{gnu-copyleft}. The central aspect of our work revolves around this type of license, due to their strength in maintaining the open-source nature of the codebase across all iterations and contributions. In light of this, we can examine and validate the condition of the data utilized for training and fine-tuning various LLMs, and compare it to the information presented in the works of these models.

\section{Approach}
The overall approach of the paper consists of two main aspects. First, we gather a comprehensive list of LLMs. We conduct a tertiary study by searching databases for recent surveys on LLMs. We then analyze the papers, repositories, and blog posts released about the identified LLMs to gain information about which datasets are used in their pretraining. Once we have collected this data from the papers, we analyze the datasets that we were able to find by seeing if they contain any copies of code that are also released under a strong copyleft license, as well as analyzing the first comment in each file to see if there are any other sources of confidential information being embedded into the weights of the models. We present a graphical summary of our approach in Figure~\ref{fig:overview}.

\subsection{Study Collection}
To understand the extent of the issue of licensed code in datasets, we must first identify which models are being trained on codebases. To understand this, we collect surveys of LLMs from online paper databases. This will give an overview of the datasets that are most commonly used in the literature and which models are trained on which datasets. We limit our search for surveys to only those that were published in 2023.

After compiling a list of LLMs from the surveys, we collect all papers, where available, and blog posts/repositories where not, that relate to the model. We further filter through these papers to identify the models that aim to only include permissively licensed code in their training procedure.
Initially, we apply a rough filter on all papers, removing all models that were not trained on code. Subsequently, we exclude all models that were not trained on file-level code. File-level code refers to code that has been extracted from repositories and not altered after collecting the data. We prioritize datasets containing file-level code as they offer a more effective means of identifying duplicated code. Extracting methods or classes may result in false positives due to common elements such as getters, setters, and common algorithms We also discard models trained exclusively on websites such as stack overflow or competition data.
After filtering the models, we first assess the availability of the datasets they were trained on. Then, we proceed to collect all publicly accessible datasets. Moreover, we extract the section from the paper/documentation that details the source of the training dataset and whether it has a specified name.
Finally, we add a class of datasets, named custom datasets. These are datasets curated by the authors of papers but not named or released. Many of these datasets are scraped from online repositories; however, not enough information is given to fully reproduce them accurately.

\begin{table}[]
  \caption{File-Level Code datasets used for training foundational models}
    \centering
    \begin{tabular}{l|l|l|r}
       \toprule 
       \multicolumn{1}{c|}{\textbf{Ref}} & \multicolumn{1}{c|}{\textbf{Dataset}} & \multicolumn{1}{c|}{\textbf{Available}} & \multicolumn{1}{c}{\textbf{Count}}\\
       \hline
       1 & Big Query  & Pay-wall &  10 \\ 
       2 & The Pile & DMCA-takedown & 12 \\ 
       3 & The Stack v1 & Open & 8 \\ 
       4 & RedPajama & Open & 3 \\ 
       5 & CodeParrot & Open & 2 \\
       6 & PaLM Dataset & Not Released & 3 \\ 
       7 & Roots & Not Open to All & 1 \\ 
       8 & SkyPile & Not Released & 1 \\
       9 & BigPython & Not Released & 2 \\
       10 & MassiveText & Not Released & 1 \\
       11 & GitHub-Code Dataset & Open & 3 \\
       12 & CodeClippy Dataset & Open & 1 \\
       13 & ExtraPythonData & Not Released & 1 \\ 
       14 & Code LLaMa Dataset & Not Released & 1 \\
       15 & Custom Dataset & Not Released & 17 \\
       \bottomrule
    \end{tabular}
    \label{tab:Datasets}
\end{table}

\subsection{Processing Collected Datasets}
After analyzing the data collected in the tertiary study, we extract a collection of datasets that are publicly available and contain at least a subset of code that has been extracted from a code base (without being modified in any way). These datasets will be the subject of the following investigations on licensed code.

We analyze two aspects of the datasets when determining whether a code file could be licensed. First, we hash all code files using the SHA-256 hash. This is a function commonly used to detect exact duplicates of code in a dataset~\cite{kocetkov2022stack}. We limit ourselves to exact duplicates, for two reasons; first, adding near deduplication adds a layer of discussion between if a piece of code is a duplicate or a slightly different, yet similar implementation of a common code structure or algorithm~\cite{bloch2022misconception}. The main goal of this paper is to analyze whether there is a reason to be concerned about licensed code in datasets, we answer this question by giving a lower bound of duplicated files, by only looking at exact copies. Second, we extract the first comment if there is any present. With the first comment, we refer to any block of comments that start in the first 20 characters of the file but may extend beyond the first 20 characters. We can search this set of comments to extract possible licenses, as well as other copyright information and disclaimers regarding the ownership and distribution of the content of the file. We release all our data in the replication package to enable future research into detecting the exact extent of license inconsistencies in datasets.

\subsection{Collect Strong Copyleft Licensed Code}
While datasets are often scraped from GitHub, some take care to only scrape permissively licensed code. In a perfect world, the overlap between code in a permissively licensed repository and a repository made available under a strong copyleft license is zero. To test how big the overlap is in the real world, and if there is a difference between datasets that check for code licenses and those that do not, we scrape our own dataset from GitHub\footnote{https://github.com}.

For generating the dataset, we query the GitHub API to generate a list of all repositories that are released under either a GPL 2.0, GPL 3.0, or AGPL license. We limit the search to 10,000 repositories and select repositories where the majority language is one of the languages we include in the investigation. We do not include files of a language when they are the minority language in a repository. In case there are less than 10,000 repositories, we use as many as available. An overview of the languages and the number of available repositories is given in ~\mbox{Table~\ref{tab:programming_languages}.} Our primary focus on these languages stemmed from their prevalence within the file-level code datasets accessible ~\mbox{to us.}

When extracting the code files from the repository, we selected files using their file extension and included all files with that extension. We did not do any secondary filtering on the length or variety of the content.
\begin{table}[]
    \caption{Programming language distribution and repository counts}
    \centering
    \begin{tabular}{p{6.4cm}|r}
        \toprule 
       \multicolumn{1}{c|}{\textbf{Programming Languages}} & \multicolumn{1}{c}{\textbf{Repositories}}\\
       \hline
         C, C\#, C++, Go, JavaScript, Java, Kotlin, Lua, Matlab, Perl, PHP, Python, R, Ruby, Rust, Shell, Swift, TypeScript & 10000 \\ 
         Assembly, Dart, Haskell & 5000 - 9999 \\
         DM, Elixir, Fortran, Julia, Lisp, OCaml, Pascal, Scala & 1000 - 4999 \\
         Agda, Erlang, SQL  & < 1000 \\ 
         \bottomrule
    \end{tabular}
    \label{tab:programming_languages}
\end{table}

\subsection{Research Questions}
Finally, we will give an overview of how we use the previously gathered information to answer each of our research questions.
\paragraph{RQ1 - Interest in Licensed Code}
To depict to what extent there is an understanding of the issue of licensed code in datasets used to train foundation models, we look at the information we gathered during the tertiary study, we use the publication date, whether the models are trained on code, and if they claim to account for permissively licensed code to extract the trends on an annual basis.

\paragraph{RQ2 - Strong Copyleft Violations}
To examine the prevalence of strong copyleft-licensed code files in datasets, we conducted two experiments.
First, we calculate the SHA-256 hash of all code files in the gathered datasets. We also calculated the SHA-256 hash of all the code files we scraped from GitHub, which only contained code from repositories with a strong copyleft license. This gives us two sets of code that we can compare for each dataset, our licensed set of hashes, and the hashes of all the files of the dataset. We report the overlap in the hashes as the number of files that appear in both the licensed repositories and the collected datasets. 
Second, we extract the first comment from each file in the collected datasets. We then search these comments for license names in order to determine whether it is referencing a GPL 2.0, GPL 3.0 or AGPL license.

\paragraph{RQ3 - Distribution disclaimers}
Finally, to evaluate whether the authors of a code file may have issues with the further distribution of their code, we apply the same procedure as for the detection of strong copyleft licenses in the first comment. However, for RQ3 we change the strings we search for to exclude all GPL 2.0, GPL 3.0, and AGPL boilerplate license declarations, and look for other language that refers to ownership and copying of code. This includes phrases as 'confidential', 'please do not share' and 'following conditions are met'. For an exhaustive list of all search strings we refer the reader to the reproduction package.

\section{Results}
To answer the research questions, we first need the results of the tertiary study. To present our results, we first give an overview of the surveys that we have collected, what papers we were able to extract from the surveys, the datasets that they resulted in, and the availability of the datasets. After collecting all the information, we proceed with the aforementioned approach to answer the research questions. 

\subsection{Tertiary Study}
For the tertiary study, we collected 6 Surveys conducted on LLMs in the year 2023, and we give an overview of these surveys in Table~\ref{tab:literature_surveys}. From these surveys, we extracted $106$ LLMs, which we further refined to 75 LLMs trained on code, 53 trained on file-level code, 23 trained on permissively licensed file-level code, and 30 trained on non-permissively licensed file-level code. This distribution can be observed in Table \ref{tab:models}. The superscripts for file-level code models denote the datasets they were trained on, based on the reference numbers assigned in Table \ref{tab:Datasets}. Due to page limitations for the references, the table is replicated with references in the reproduction package.

There were a total of 14 datasets that we identified. Of these datasets, 7 were not released by the authors, 5 were fully open, 1 was released selectively to practitioners that applied, 1 was open to all but required payment to generate and download, and 1 was removed due to a DMCA takedown request, but re-uploaded by a third party without the offending material.

Aside from these datasets, some models were trained on scrapes of GitHub that the authors conducted themselves. Often, there was not enough information presented to fully replicate the dataset the authors claimed to have created. These cases were not considered in this work.

The final set of datasets used for further investigations are \textit{The Pile}\footnote{https://huggingface.co/datasets/monology/pile-uncopyrighted}, \textit{The Stack v1}\footnote{https://huggingface.co/datasets/bigcode/the-stack}, \textit{RedPajama}\footnote{https://huggingface.co/datasets/togethercomputer/RedPajama-Data-1T}, \textit{CodeParrot}\footnote{https://huggingface.co/datasets/codeparrot/codeparrot-clean-valid}, \textit{Github-Code}\footnote{https://huggingface.co/datasets/codeparrot/github-code}, \textit{CodeClippy}\footnote{https://huggingface.co/datasets/CodedotAI/code\_clippy\_github}.

 Among these datasets, \textit{The Stack v1}, \textit{CodeParrot}, \textit{GitHub-Code}, and \textit{RedPajama} incorporate a license field within their data structure. Additionally, permissive code is mentioned in reference to \textit{The Stack v1} and \textit{RedPajama}, whereas \textit{CodeParrot} and \textit{GitHub-Code} incorporate both permissive and copyleft licenses in their code. In the absence of a license field or a specific mention of permissive code for \textit{CodeClippy} and \textit{The Pile}, an examination of their data reveals a mixture of both permissive and copyleft licenses.

\subsection{RQ1 - Interest in Licensed Code}
\begin{figure}[]
    \centering
    \includegraphics[scale=.55]{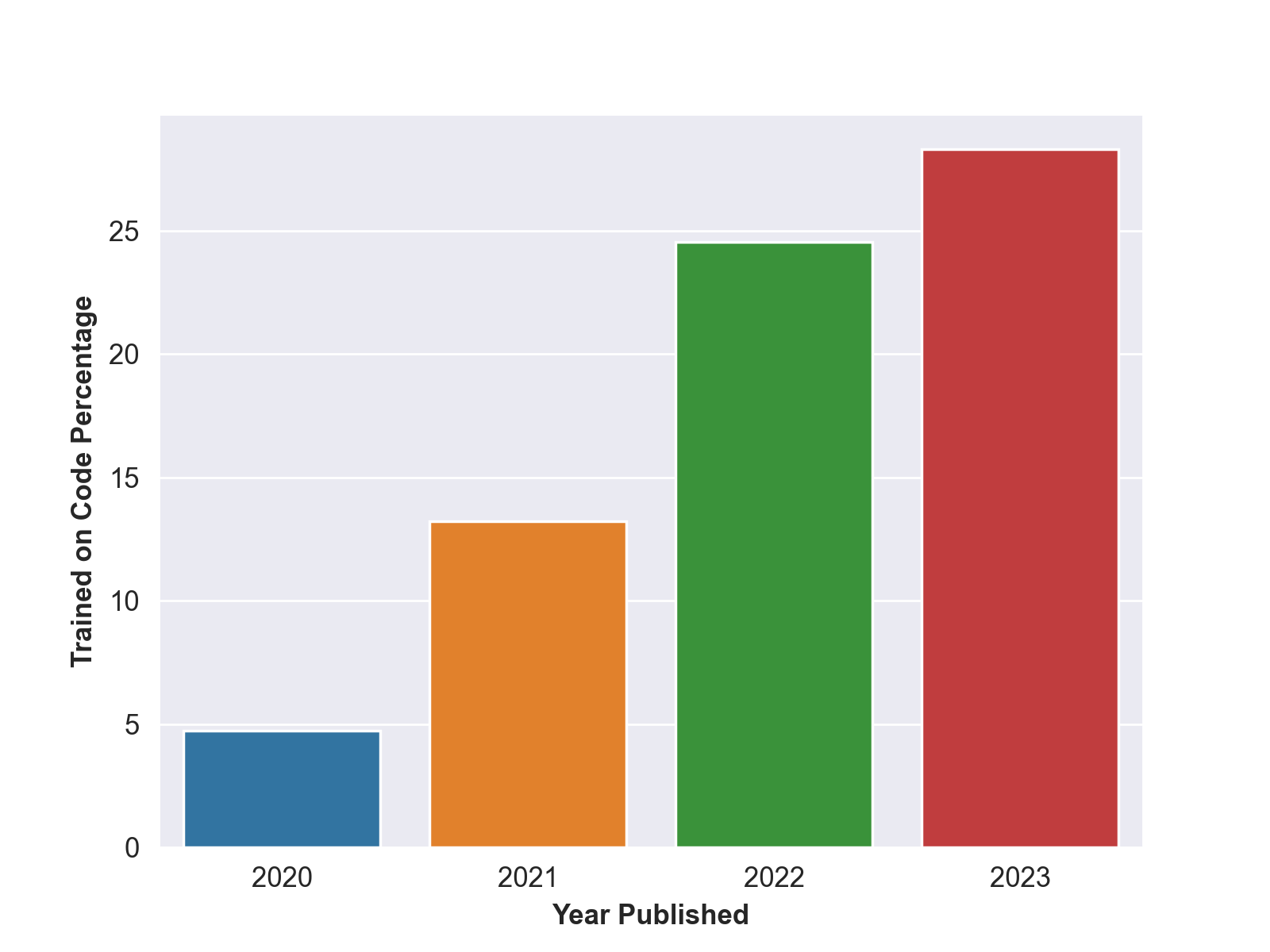}
    \caption{Percentage of LLMs trained on code per year over the total number of LLMs}
    \label{fig:trained_code}
\end{figure}

\begin{figure}[]
    \centering
    \includegraphics[scale=.55]{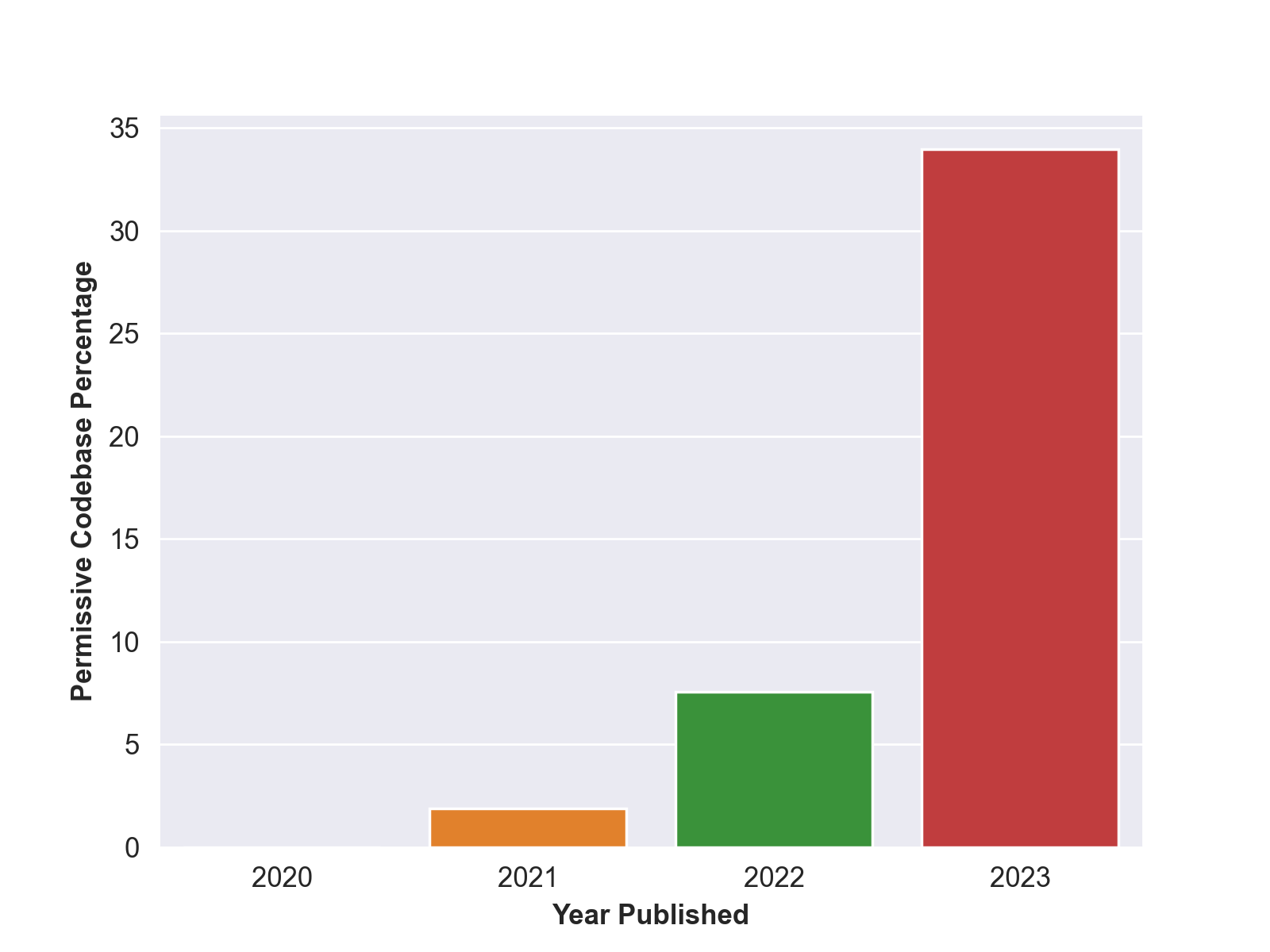}
    \caption{Percentage of LLMs trained on permissive file code per year over the total number of LLMs trained on file level code}
    \label{fig:permissive_code}
\end{figure}

To judge the interest of the field on the presence of licensed code in training data, we analyse both the presence of code in training setups, as well as the references to code licenses in papers. We show the results of this in Figure~\ref{fig:trained_code} and Figure~\ref{fig:permissive_code}.

We see that there has been a gradual increase in the use of code in training setups since 2020. We attribute this rise in interest to the popularity of products such as copilot\footnote{https://github.com/features/copilot} and ChatGPT\footnote{https://openai.com/chatgpt}, as well as the emerging benefit of using code data when training models for natural language reasoning~\cite{anonymous2023at}.

We further evaluate the attention being paid to permissively licensed code when training models. We see that there has been a large increase in interest in using datasets that only contain permissively licensed code. Looking at Figure~\ref{fig:permissive_code} we see that there is a large jump in papers referencing permissively licensed code. We attribute this to the recent increase in news coverage of other generative models, such as stable diffusion~\cite{rombach2021highresolution} and datasets such as \textit{The Pile} getting targeted by legal action~\cite{gao2020pile}. Furthermore, this increase in interest also coincides with the release of \textit{The Stack v1}~\cite{kocetkov2022stack}, one of the datasets under investigation, which puts a large emphasis on containing only permissively licensed code.

\subsection{RQ2 - Strong Copyleft Inconsistencies}
To answer RQ2 we will approach the problem from two directions. First, we gather a dataset that we scraped from GitHub containing only repositories with strong copyleft licenses. We then check to see if there are exact copies of these files in the datasets used for training. Additionally, we retrieve the first comment from all files and search for substrings that match the boilerplate license disclaimer commonly used by strong copyleft licenses.

Table \ref{tab:CopiedCode} features data on both the number of exact duplicates (using the SHA-256 hash) and the number of comments denoting the use of strong copyleft licenses across all datasets. The dotted line separates the datasets that claim to exclusively utilize permissive licenses, positioned at the top, from the remaining ones. After analyzing the obtained results, it is evident that there is a considerable overlap of exact duplicates between the dataset containing only strong copyleft licensed repositories and other datasets. 

Although we see that the percentage of files that overlap is lower for \textit{The Stack v1} and \textit{RedPajama} than it is for other datasets, such as \textit{The Pile}, \textit{CodeParrot}, and \textit{CodeClippy} which do not filter on licenses, the datasets that checked for licenses still have a larger overlap than \textit{Github-Code}, which did not check for licenses.

Furthermore, we see that when looking at comments that contain the license of the file, we see that the datasets that check the repositories for licenses have a significantly smaller match percentage with strong copyleft licenses than those without.
\begin{table*}[]
    \caption{Amount of code files found to be associated with a strong copyleft license}
    \centering
    \begin{tabular}{l|r|r|r|r|r}
        \toprule 
        \multicolumn{1}{c|}{\multirow{2}{*}{\textbf{Dataset}}} & \multicolumn{1}{c|}{\multirow{2}{*}{\textbf{Files}}} & \multicolumn{2}{c|}{\textbf{Exact Duplicates}} & \multicolumn{2}{c}{\textbf{License Comments}} \\
        \cline{3-6}
        & & \multicolumn{1}{c|}{\textbf{Count}} & \multicolumn{1}{c|}{\textbf{Percentage}} & \multicolumn{1}{c|}{\textbf{Count}} & \multicolumn{1}{c}{\textbf{Percentage}} \\
        \hline
        The Stack v1 & 262,678,972 & 16,122,976 & 6.14\%& 2,067,830 & 0.78\%\\
        RedPajama & 28,793,312 & 1,579,521 & 5.49\%& 15,544 & 0.05\%\\
        \hdashline
        The Pile & 18,044,000 & 4,113,263 & 22.80\%& 823,546 & 4.56\%\\
        CodeParrot & 18,695,559 & 2,681,590 & 14.34\%& 2,844,150 & 15.21\%\\
        GitHub-Code & 115,086,922 & 5,537,734 & 4.81\%& 7,548,615 & 6.56\%\\
        CodeClippy & 71,140,482 & 7,993,768 & 11.24\%& 2,823,923 & 3.97\%\\
        \bottomrule
    \end{tabular}
    \label{tab:CopiedCode}
\end{table*}

\subsection{RQ3 - Distribution Disclaimers}
Finally, we analyze the prevalence of any language in the opening comment of all files. The goal of this is to identify if any comments show the authors/owners of a file did not want the contents to be shared or distributed. Table~\ref{tab:Confidential_code} presents the results of the searches, detailing the number of copyright disclaimers found in the first comments and the number of first comments across datasets. Additionally, we provide such an example in Figure~\ref{fig:example_comment} to illustrate the nature of the identified comments.

We see that when looking a the percentages of comments containing distribution disclaimers to total amount of first comments, the prevalence of distribution disclaimers is between $5\%$ and $7.5\%$ for all datasets except for \textit{RedPajama} that had $1.3\%$. This shows that there are a large number of code files that also have some restrictions on redistribution; however, do not use the standard disclaimer or a specific license.

\begin{table}[]
    \centering
    \caption{Amount of code files found to be associated with some form of ownership/copyright disclaimer}
    \begin{tabular}{l|r|r|r}
        \toprule 
        \multirow{2}{*}{\textbf{Dataset}} & \multicolumn{2}{c|}{\textbf{Copyright}} & \multirow{2}{*}{\textbf{First Comments}}\\
        \cline{2-3}
        & \textbf{Count} & \textbf{Percentage} & \\
        \hline
        The Stack v1 & 5,073,823 & 6.54\% & 77,595,559\\
        RedPajama &  30,500 & 1.34\%& 2,281,378\\
        \hdashline
        ThePile & 501,877 & 7.39\% & 6,794,995\\
        CodeParrot & 773,062 & 5.38\% & 14,372,397\\
        GitHub-Code & 2,669,845 & 5.89\% & 45,301,797 \\
        CodeClippy & 1,695,556 & 6.72\% & 25,223,157\\
        \bottomrule
    \end{tabular}
    \label{tab:Confidential_code}
\end{table}

\begin{figure}
\centering
    \begin{minted}[
    linenos,frame=single,
    breaklines=true,
    framerule=0.5pt,
    breaksymbolleft=,
    numbersep=1mm,
    fontsize=\footnotesize]{text}
    <Company> all rights reserved.
    this software contains proprietary and confidential
    information of <Company> and its contributors.
    use, disclosure and reproduction is prohibited without
    prior consent.
    \end{minted}
    \caption{Restrictions on sharing and distributing code contained in a file, extracted from the \textit{RedPajama} dataset}
    \label{fig:example_comment}
\end{figure}

\section{Discussion}
In the results, we found interesting trends regarding the popularity of permissive code, the prevalence of incorrectly classified licensed code, and the presence of other language that places limitations on the distribution and use of code files. We begin by demonstrating how the findings relate to the wider field of LLMs for code; we then give some recommendations to other practitioners about how to deal with code licenses. We complete the section with an overview of avenues for future work and any limitations we found with our approach.

\subsection{Implications}
The real-world implications of our study affect a number different stakeholders.  For maintainers and curators of datasets, we have shown that there is more information present in code comments in relation to the distribution of code than only the license. We have also shown that there is a significant overlap in repositories that have different licenses, making it hard to judge where code originates from and what license should be applied to it. This also affects practitioners who train LLMs, as they carry the risk of their models being targeted if they are trained on licensed code. Finally, end users of LLMs need to be wary of accidentally inserting licensed code into their code bases as the final outcome of doing so with strong copyleft code is not yet known but could in theory lead to the open-sourcing of their code bases. The main question that these implications raise for the wider field of LLMs for code is who is responsible for the training data and output of the final models.

\subsection{Future Work}
From the results gathered, we have identified a number of venues for future work. The main areas for future work relate to the analysis of comments and extraction of licenses/intent from comments, as well as determining if a piece of code is licensed and under which license it is licensed.

\paragraph{Code Comment Intent}
We have seen that aside from the legal aspects of potential license violations, there is also an issue with the consent of the authors to include code in training data. Although some curators provide an opt-out system to be removed from the dataset~\cite{kocetkov2022stack}, it should be possible to detect an authors intent from the comment, if present.

\paragraph{Duplicate Code}
Similar to using basic strings to search for comment intent, we also used a basic exact duplicate finding technique using the SHA-256 hash function. Although it showed the presence of duplicates between our two sets of data, it takes minimal effort to avoid detection. Small code changes, which could be automatic, such as changing from 4 space indent to 2 space indent, would give a false negative. Furthermore, removing licensing information would also lead to an undetected duplicate. In future work, we believe that it would be beneficial to the community to create a taxonomy of code changes that take place when code is copied from one repository to another. In addition, it gives a better overview of how much code of each category is copied. This would give a good understanding of potential issues with copied code and is a good way to relate it to the legal implications of fair use.

\paragraph{Identifying a License Violation}
One of the issues with only detecting duplicates is that we do not know for certain if a license was violated by including it in a non-permissive dataset vs. a permissive dataset as we only know that the file is duplicated. An interesting area of research would be to analyze networks of code repositories, such as GitHub, to automatically detect whether a file or snippet of code was copied in a way that violates the license. This can be a quick analysis of license changes when looking at forks, or a more thorough investigation of tracking code changes and different versions of code through time and different repositories.

It must also be said that this problem is not unique to LLM datasets. Previous work has focused on the provenance of code~\cite{davies2011software} and the identification of license violations in large software projects~\cite{german2009license}, as well as on the automatic identification of licenses for code files in software projects~\cite{german2010sentence}. 

\subsection{Recommendations}
To address the issues uncovered during our investigation, we propose several recommendations for developers who are involved in training LLMs.

First and foremost, our investigation has revealed that no LLM to date has been trained on a dataset entirely devoid of licensing issues. Given that fine-tuning a model retains the information embedded in its weights, it is crucial to be mindful of this when choosing a model for the fine-tuning process.

Moreover, our investigation has revealed that even in cases where providers of permissively licensed source code datasets have good intentions, straightforward searches often result in numerous inconsistencies. Therefore, we recommend that when fine-tuning or training a model from scratch, it is advisable to conduct thorough searches of the datasets to identify and rectify any potential licensing issues before starting training. This precaution is particularly important, as removing information from the model's weights has not been proven to be a foolproof solution~\cite{wu2023depn}.

Finally, we recommend adapting and scaling up existing methods for license detection to be able to work for large datasets of code in order to reduce the number of license inconsistencies that could be present in datasets scraped from online repositories.

\subsection{Limitations}
Our approach faces two primary limitations. Firstly, the dynamic nature of code means that copied snippets are subject to change over time. Secondly, there is a notable lack of transparency from many entities regarding their training methodologies. We elaborate on these limitations in the subsequent sections of this paper.

\paragraph{Collected code}
As codebases are in a constant state of evolution, we have established a fixed point in time for gathering code. However, obtaining copies of our strongly copyleft-licensed dataset from the exact moment each individual dataset was curated proves challenging due to resource limitations. In some cases, it is simply impossible as the specific scraping date of these datasets remains undisclosed.
Furthermore, since the creation of our dataset, changes may have occurred. Some repositories may have been deleted or converted to private status, while new repositories have emerged. Consequently, our datasets may contain slightly different sets of code. Nevertheless, it is worth noting that our primary focus is on identifying the presence of strong copyright-protected code within these datasets. The discrepancies in dataset composition do not impact the conclusions drawn in this paper.

\paragraph{Non-reproducible Papers}
Finally, one of the limitations we experienced when collecting information about the training of proprietary models, and analysis of their datasets is that some companies were not transparent about how the models were trained. In many cases when a custom dataset was scraped, information that would be needed for an exact replication of the dataset was missing. This was usually information such as the date it was scraped, which exact repositories were scraped, or if there were any additional filtering criteria for the code files. Furthermore, we noticed that, especially when companies described their models in blog posts or white papers, they were very ambiguous about what exact data the models were trained on. Oftentimes, the exact nature of the data could not be inferred from the paper. Unfortunately, due to the competitive nature of LLMs and the price associated with curating data and training models, this trend will probably continue.

\section{Conclusion}
We evaluated the presence of code licensed under a strong copyleft license in datasets used to train foundation models. To analyze the datasets used and the attention paid to permissive code, we collected 6 literature surveys, from which we extracted 106 foundation models. Of these models, $53$ were trained on file-level code and $23$ specifically referenced training exclusively on permissively licensed code. We further analyzed the papers and extracted $30$ datasets that worked with file-level code. Of these datasets, $17$ were custom GitHub scrapes. Furthermore, of these datasets, we were able to access 6 that were distributed online without restrictions. 
Of these 6 datasets, 4 did not filter the collected code on licenses while the other 2 did. In total, we collected 514 million files of code across all 6 datasets that we used to evaluate the presence of strong copyleft licenses.

To evaluate the presence of strong copyleft-licensed code we scraped GitHub, we selected up to 10,000 repositories released under either a GPL 2.0, GPL 3.0, or AGLP license, covering 32 languages. This resulted in a dataset of 35 million code files.
We calculated the SHA-256 hash of all collected datasets and the strong copyleft dataset we collected ourselves. We then use the hashes to look for an overlap of exact copies. We found that all datasets had a substantial overlap with the dataset of strong copyleft licensed code. Although datasets that checked for permissively licensed code generally had less of an overlap, there was still an overlap of at least $5\%$.

Furthermore, we extracted the first comments from the datasets to detect licenses that are used on a file level. We saw that when detecting license comments, datasets that claim to contain permissively licensed code perform better, with $0.05\%$ and $0.8\%$ of files having a license comment; however, this still amounts to more than 2 million files with a strong copyleft license in the case of \textit{The Stack v1}.
Finally, we also analyzed the comments gathered from the datasets, to evaluate the presence of any other disclaimers concerning copying the contents, which is not directly related to any license. We find that there is a higher prevalence of these non-license disclaimers than there are comments containing a strong copyleft license. These disclaimers are also prevalent in all datasets. 

To enable further investigations into detecting and removing licensed code from public datasets, we release a dataset containing all comments that appear at the start of a file of code that we collected during this investigation. The 450 million files resulted in a dataset of 171 million code comments, and we removed all PII before releasing the data.
Overall we have shown that while the interest in creating datasets that contain only permissive code has grown rapidly in the last year. 
However, there is evidence of license inconsistencies that need to be addressed in order to fully avoid future problems with regards to licensing.

\section{Data availability}
To make our experiments reproducible, we released a replication package at \url{www.github.com/AISE-TUDelft/CodeLicensingExploration}. We share the repositories we collected, the raw results of the tertiary study, and the code we used. We also upload the dataset containing the leading comments to huggingface at \url{www.huggingface.co/datasets/AISE-TUDelft/leading-comments}, we used the StarPII\footnote{https://huggingface.co/bigcode/starpii} model to remove any Personal Identifiable Information (PII) from the dataset prior to uploading. 

\bibliographystyle{ACM-Reference-Format}
\bibliography{main}

\end{document}